\documentclass[conference, 10pt]{IEEEtran}

\usepackage[cmex10]{amsmath}
\usepackage{url}
\usepackage{graphicx}
\usepackage{color}
\usepackage{placeins}
\usepackage{float}
\usepackage{tabularx,colortbl}
\usepackage{xfrac}

\begin{document}

\title{Vulnerability of LTE to Hostile Interference}

\IEEEoverridecommandlockouts

\author{
    \begin{tabular}{ccccccc}
      {\bf Marc Lichtman}$^1$   &~&  {\bf Jeffrey H. Reed}$^1$   &~&  {\bf T. Charles Clancy}$^2$   &~&   {\bf Mark Norton}$^3$ \\
      marcll@vt.edu &~& reedjh@vt.edu &~&  tcc@vt.edu &~& mark.norton@osd.mil   \\
    \end{tabular}\\~\\
    $^1$Wireless @ Virginia Tech, Virginia Tech, Blacksburg, VA\\
    $^2$Hume Center for National Security and Technology, Virginia Tech, Arlington, VA \\
    $^3$Office of the Chief Information Officer, Defense Pentagon, Washington, DC
    \thanks{
      The views expressed in the article are views of the
      authors and do not necessarily reflect the official policy or
      position of the Department of Defense or United States
      Government.
    }
}


\maketitle

\begin{abstract}
LTE is well on its way to becoming the primary cellular standard, due to its performance and low cost.  Over the next decade we will become dependent on LTE, which is why we must ensure it is secure and available when we need it.  Unfortunately, like any wireless technology, disruption through radio jamming is possible.  This paper investigates the extent to which LTE is vulnerable to intentional jamming, by analyzing the components of the LTE downlink and uplink signals.  The LTE physical layer consists of several physical channels and signals, most of which are vital to the operation of the link.  By taking into account the density of these physical channels and signals with respect to the entire frame, as well as the modulation and coding schemes involved, we come up with a series of vulnerability metrics in the form of jammer to signal ratios.  The ``weakest links'' of the LTE signals are then identified, and used to establish the overall vulnerability of LTE to hostile interference.
\end{abstract}

\begin{IEEEkeywords} LTE, LTE security, jamming, interference \end{IEEEkeywords}

\section{Introduction}
LTE service is widely available in the United States and other countries around the world.  LTE is well on its way to becoming the primary cellular standard, due to its performance and low cost.  In addition to everyday use, cellular networks are often used to broadcast emergency information during natural disasters and other crises.  Over the next decade we will become dependent on LTE, which is why we must ensure it is secure and available when we need it.  Unfortunately, like any wireless technology, disruption through radio jamming is possible.

The objective of this paper is to analyze the extent to which LTE is vulnerable to jamming, and to derive metrics in order to compare possible weak points in the downlink and uplink signals.  In order to derive metrics that represent the effectiveness of different jamming techniques, we will introduce two different symbols corresponding to the jammer-to-signal ratio (J/S). ${J/S}_{RE}$ will correspond to a J/S when only taking into account the specific subcarriers and OFDM symbols (a.k.a. resource elements) being jammed.  J/S averaged over an entire frame will be referred to as ${J/S}_F$.  We will limit the scope of this paper by only analyzing Frequency Division Duplex (FDD) configured LTE, due to its widespread use.  However, much of the analysis included in this paper can be applied to Time Division Duplex (TDD) as well.

Attacks on LTE can be grouped into two broad categories; Denial of Service (DOS) and information extracting.  Jamming attacks are typically used to cause DOS, while the area of cyber security deals with attacks that  extract information, cause DOS, or both. There is very little openly available literature related to attacks on LTE.  The authors of \cite{6154001} introduce the non-access-stratum request attack, which causes DOS by flooding the Home Subscriber Server (HSS). An attack designed to cause degradation of service is described in \cite{4394792}, in which an attacker sends fake buffer status reports to the eNodeB, which causes the eNodeB to assign excessive resources to users which don't exist. The author of \cite{5962467} analyzes OFDM denial using barrage jamming, pilot tone jamming, and pilot tone nulling.  An overview of the security of LTE availability is given in \cite{joversecurity}.

\section{Background of LTE}

\begin{figure}[htb]
  \centering
  \includegraphics[trim = 0mm 4mm 0mm 0mm, clip, width=2in]{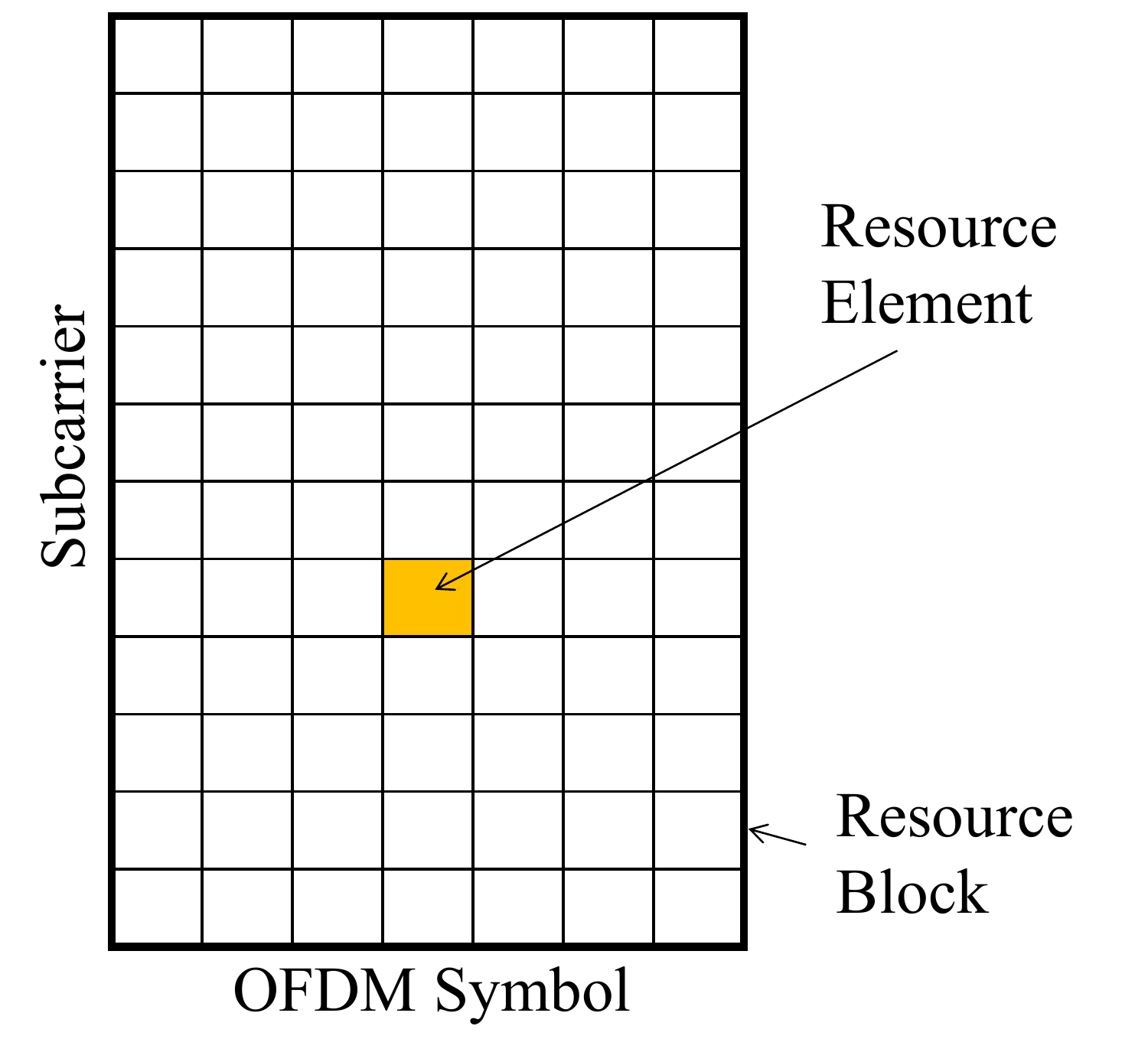}
  \caption{A Single LTE Resource Block}\label{resource_block}
\end{figure}

Orthogonal Frequency-Division Multiple Access (OFDMA) is the multiple access scheme used in the LTE downlink \cite{phy_channels_modulation}.  OFDMA uses multiple carriers, which makes it effective in a frequency selective channel.  Each subcarrier carries a separate stream of information, causing information to be mapped in both the time and frequency domain.  This leads to the Orthogonal Frequency-Division Multiplexing (OFDM) time-frequency lattice, which is a two-dimensional grid used to represent how information is mapped to both the subcarrier and the OFDM symbol.  In LTE, one subcarrier over one OFDM symbol is called a resource element, and 12 consecutive subcarriers over 7 OFDM symbols are combined to form a resource block, as shown in Figure \ref{resource_block}. This method of information mapping allows a jammer to selectively jam information in both the time and frequency domain.
Single Carrier-Frequency Division Multiple Access (SC-FDMA) is the multiple-access scheme selected for the LTE uplink \cite{phy_channels_modulation}. LTE user devices are known as User Equipment (UE). The UE accesses the LTE network by connecting to the eNodeB, which acts as a base station.

\section{Vulnerability of LTE Physical Channels}
The following subsections investigate the various LTE physical channels with the goal of finding the minimum J/S required to cause the physical channel to be corrupted beyond functionality.  The J/S thresholds associated with the physical channels are largely based on the modulation and coding scheme used in each channel, and in each subsection we approximate the Bit Error Rate (BER) or Block Error Rate (BLER) required to cause a corrupt channel. The actual BER/BLER threshold is based off of numerous factors on many different layers, and would be best acquired empirically. Table \ref{phy_channels} highlights the parameters associated with each physical channel \cite{phy_channels_modulation,multiplexing_and_coding}.

\begin{table}[b]
    \centering
        \caption{Physical Channel Modulation and Coding Schemes}\label{phy_channels}
        \begin{tabular}{| l | l | l | l |}
            \hline
            \textbf{Channel} & \textbf{Modulation} & \textbf{Coding} & \textbf{Coding Rate} \\
            \hline
            PDSCH & \{4,16,64\}-QAM & Turbo & Adaptive \\
            \hline
            PBCH & QPSK & Convolutional & 1/48 \\
            \hline
            PCFICH & QPSK & Block & 1/16 \\
            \hline
            PDCCH & QPSK & Convolutional & 1/3 \\
            \hline
            PHICH & BPSK & Repetition & 1/3 \\
            \hline
            PUSCH & \{4,16,64\}-QAM & Turbo & Adaptive\\
            \hline
            PUCCH & BPSK, QPSK & Convolutional & 1/3 \\
            \hline
            PRACH & ZC Sequences & N/A & N/A \\
            \hline
        \end{tabular}
\end{table}

\subsection{PDSCH and PUSCH (User Data)}
The Physical Downlink Shared Channel (PDSCH) and Physical Uplink Shared Channel (PUSCH) are used to transmit user data to and from the eNodeB.  These two channels utilize adaptive modulation and coding, and undergo either QPSK, 16-QAM, or 64-QAM depending on the channel quality \cite{phy_channels_modulation}.  Both channels use turbo coding for forward error correction, with a coding rate as low as 0.076 (when using rate-matching)  \cite{phy_procedures}.  In the presence of an interferer, we will assume that the modulation ratchets down to QPSK at 0.076 rate coding.  The authors of \cite{6240303} analyze low-rate turbo codes using OFDM and SC-FDMA in a typical urban channel.  At a coding rate of $\sfrac{1}{15}$, an average SNR of around -7 dB results in a BLER of 0.1 \cite{6240303}.  Determining the exact effects of a 0.1 BLER on the PDSCH and PUSCH is beyond the scope of this paper; however it is predicted that this BLER will lead to an overwhelming amount of retransmissions.  We will therefore estimate the ${J/S}_{RE}$ threshold for these physical channels to be 7 dB.

\subsection{PCFICH (Downlink Control Format Indicator)}
The Physical Control Format Indicator Channel (PCFICH) is used to send the UE information regarding where the Physical Downlink Control Channel (PDCCH) is located in the time-frequency lattice.  Without successful decoding of this information, the UE will not be able to decode the PDCCH.  The PDCCH contains information regarding UE resource allocation, which is vital to the LTE service.  Although it is possible to jam the PDCCH directly, we will first analyze the J/S threshold of the PCFICH.

\begin{figure*}
  \centering
  \includegraphics[trim = 0mm 5mm 0mm 0mm, clip, width=7.1in]{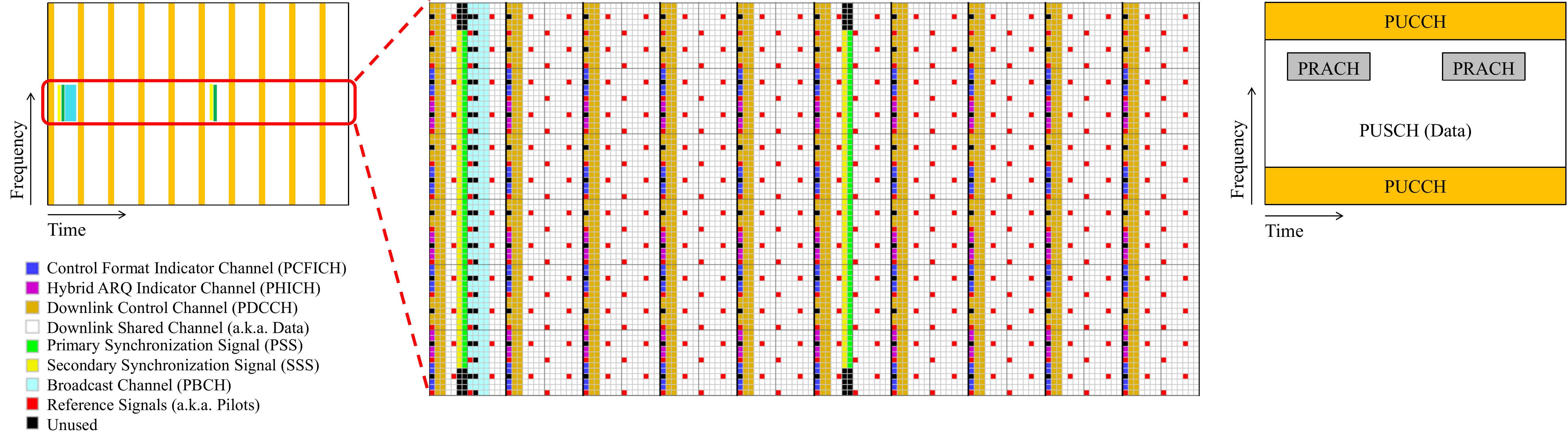}
  \caption{LTE Downlink Frame (Left/Center) and Uplink Frame (Right)}
  \label{lattice}
\end{figure*}

The PCFICH appears in only one symbol per subframe, and occupies 16 subcarriers.  Jamming the PCFICH consists of transmitting on top of the 16 subcarriers.  The location of the 16 subcarriers is not static; it is determined by the eNodeB's complete cell ID \cite{phy_channels_modulation}.  This ID is carried in the PSS and SSS, and therefore selectively jamming the PCFICH requires the jammer to synchronize to both downlink synchronization signals.  This also limits a PCFICH jamming attack to a single cell.  Figure \ref{lattice} includes an example FDD downlink frame (left and center graphics), by displaying the color-coded time-frequency lattice.  The resource elements used for the PCFICH are shown in blue.

The information carried on the PCFICH is a two bit indicator, which is encoded using a block code of rate $\sfrac{1}{16}$.  Due to the function of the PCFICH, successful jamming requires transmitting at a high enough power to cause a BER near 0.5. The authors of \cite{Proakis} use BPSK in an Additive White Gaussian Noise (AWGN) channel to show that a soft-decision decoded block code with rate $\sfrac{1}{16}$ can be decoded at an SNR down to -1.5 dB.

\subsection{PUCCH (Uplink Control Channel)}
The Physical Uplink Control Channel (PUCCH) is used to send the eNodeB a variety of control information, including scheduling requests, Hybrid Automatic Repeat Request (HARQ) acknowledgements, and channel quality indicators.  The PUCCH is mapped to the resource blocks on the edges of the system bandwidth, as shown in Figure \ref{lattice}.  This allows PUCCH jamming to be possible when the only a priori knowledge is the LTE system bandwidth and center frequency.  For an uplink BW of 10 MHz, roughly 16 resource blocks (or 192 subcarriers) are allocated to the PUCCH \cite{phy_channels_modulation}.  Therefore, PUCCH jamming requires jamming about 25\% - 30\% of the uplink system bandwidth.  The PUCCH is modulated with a mix of BPSK and QPSK, and uses $\sfrac{1}{3}$ rate convolutional coding.  It can be shown that BPSK under an AWGN channel reaches a BER of 0.1 at around 2 dB of SNR, when using soft-decision decoding at rate $\sfrac{1}{3}$ \cite{771235}.

\subsection{PBCH (Downlink Broadcast Channel)}
After synchronizing with the PSS and SSS, the UE receives more information about the cell by decoding the Master Information Block (MIB), which is transmitted on the Physical Broadcast Channel (PBCH).  The MIB contains information essential for initial access to a cell \cite{LTEch9}.  It consists of 14 bits that contain the downlink system bandwidth, the PHICH size, and information allowing frame synchronization.  It is mapped to the center 72 subcarriers, and appears in the first subframe of every frame.  The PBCH is transmitted using QPSK, and uses a 16-bit CRC as a form of error detection.  It also uses a special channel coding scheme that creates four individually self-decodable units, each with rate $\sfrac{1}{12}$, but all four units can be decoded together for a coding rate of $\sfrac{1}{48}$.  This is accomplished using a mixture of repetition coding and convolutional coding. An uncoded QPSK signal reaches a BER of 0.1 at roughly 0 dB of SNR in an AWGN channel, and the PBCH coding scheme does not provide a significant gain at this value of SNR \cite{Proakis}.  We will therefore estimate the ${J/S}_{RE}$ threshold for the PBCH to be 0 dB.

\subsection{PHICH (Hybrid-ARQ Indicator Channel) }
Downlink acknowledgements (ACK/NACK) are sent on the Physical Hybrid-ARQ Indicator Channel (PHICH).  The PHICH uses BPSK with repetition-3 coding \cite{phy_channels_modulation}.  BPSK using repetition-3 coding in a fading channel reaches a BER of 0.1 at about 2 dB of SNR \cite{Proakis}, leading to a ${J/S}_{RE}$ threshold of roughly -2 dB.

\section{Vulnerability of LTE Physical Layer Signals}
\subsection{Primary and Secondary Synchronization Signals}
Detecting the PSS is the first step a UE takes in accessing a cell.  The PSS signal is constructed from a Zadoff-Chu (ZC) sequence, which are complex-valued sequences that have constant amplitude.  An odd-length ZC sequence is given by
\begin{equation}
x_q[k]=\exp \left[ -j\frac{\pi qk(k+1)}{N} \right]
\end{equation}
where $N$ is the length of the sequence and $q$ is the ZC sequence root index \cite{LTEch7}.  The PSS uses a sequence length of 63, and there are three PSS sequences used in LTE, usually corresponding to one of three sectors.

A jamming attack against the PSS would require a fairly high ${J/S}_{RE}$, because the PSS is designed to be detected at high interference levels, so that the UE can also detect neighboring cells.  A more effective method of corrupting the PSS would be to simply transmit all three PSS sequences, thus spoofing the synchronization signal.  If the jammer's received power at the UE is greater than the eNodeB's (a ${J/S}_{RE}$ over 0 dB), then the UE is most likely going to synchronize to the bogus PSS.  The actual detection algorithm is left up to the vendor, so we will assume that a ${J/S}_{RE}$ of 3 dB is enough to cause the PSS subsystem to fail nearly all of the time.  PSS spoofing will not immediately cause DOS; it will prevent new UEs from accessing the cell(s) and cause UEs in idle mode to reselect a bogus cell. While it is possible for the blacklisting mechanism implemented in the UE to effectively ignore the bogus signals, we will assume the system gets ``confused'' when it detects a valid PSS with no associated SSS and does not know how to deal with it.

Detecting the SSS is the second required step in accessing an LTE cell.  The SSS provides frame timing information, cell group ID, cyclic prefix length, and TDD/FDD configuration.  It is a BPSK modulated signal made up of m-sequences.  Simply jamming the SSS is not effective for the same reason as jamming the PSS, and spoofing the SSS requires synchronizing with the target cell, because the UE expects the SSS to be in a certain location.  For these reasons, SSS jamming and spoofing is left out of the final comparison.

\begin{table*}
  \centering
  \caption{Comparison of Vulnerabilities}\label{comparison_table}%
    \begin{tabular}{|l|l|l|l|l|l|l|l|}
    \hline
    \textbf{Jamming Method} & \textbf{\# of REs} & \textbf{\% of REs} & $\boldsymbol{{J/S}_{RE}}$ & \textbf{Synch. Required} & \textbf{Complexity} & $\boldsymbol{{J/S}_{F}}$\\
    \hline
    Barrage Jamming                     & 84000   & 100\%  & -2 dB   & No  & Very Low & -2 dB\\
    \hline
    RS Jamming                          & 4000    & 5\%    & 4 dB   & Yes & High  & -9 dB\\
    \hline
    Center 6 Resource Blocks Jamming    & 10080   & 12\%   & -2 dB  & No  & Very Low & -11 dB\\
    \hline
    PSS Spoofing                        & 378     & 0.45\% & 3 dB   & No  & Medium & -20 dB\\
    \hline
    PCFICH Jamming                      & 160     & 0.2\%  & 1.5 dB & Yes & High  & -25 dB\\
    \hline
    PUCCH Jamming                       & 21000   & 25\%   & -2 dB  & No  & Low   & -8 dB\\
    \hline
    \end{tabular}%

\end{table*}

\subsection{Downlink Reference Signals}
In order for an OFDM receiver to estimate the channel and perform frequency-domain equalization, there must be known symbols periodically transmitted.  Although these known symbols are often referred to as pilots or reference symbols, in the LTE specification they are called Reference Signals (RSs).  In the downlink, RSs are multiplexed in both time and frequency, as shown in Figure \ref{lattice}. RSs occupy roughly 14\% of the resource elements in a frame.  The location of the RSs in time and frequency is based on the cell ID.  All RSs are QPSK modulated, and use a length-31 Gold sequence that is initialized with a value based on the cell ID.

It is shown in \cite{1311315} and \cite{5962467} that jamming a subcarrier that contains RSs leads to a higher BER than one that only contains data.  A RS jamming attack requires detecting the target eNodeB's PSS and SSS in order to retrieve the cell ID.  However, it does not require estimating the channels involved, due to the long symbol duration (71 microseconds).  Even if there is 5 miles between the jammer and UE, there would only be a propagation delay of 27 microseconds.  To compensate for this delay, the jammer would only have to start transmitting a fraction of a symbol early.

A RS jamming simulation performed in \cite{5962467} shows that when using a QPSK signal and an SNR of 10 dB, an overall J/S (${J/S}_F$) of -5 dB causes a BER of 0.1.  The simulated system uses a pilot density of 1/8, resulting in a J/S of 4 dB when only considering the jammed resource elements (i.e. the value of ${J/S}_{RE}$).  An 8-tap channel is used, and the signal is generated with a cyclic prefix length of 1/8 and a 256-point FFT.  Although the simulated system is not exactly the same as LTE, it provides a reasonable approximation of ${J/S}_{RE}$.

\section{Comparison of Vulnerabilities}
In order to compare the extent of each vulnerability, we use a system bandwidth of 10 MHz (this equates to 50 resource blocks).  Table \ref{comparison_table} lists the various forms of jamming, along with key metrics.  Jamming techniques that do not perform better than barrage jamming are left off this table.  Note that jamming the center 6 resource blocks means jamming the PBCH without time synchronization, using a 100\% duty cycle waveform.

The second and third columns of Table \ref{comparison_table} show the number and percent of resource elements per frame that must be jammed in each attack.  The minimum ${J/S}_{RE}$ column shows the ${J/S}_{RE}$ required to cause immediate denial of the channel or signal, which is a value estimated in the analysis of each section, and is meant to be a rough estimate.  Two of the attacks require the jammer to maintain time domain synchronization with target cells, using the PSS and SSS.  The complexity metric is based on the amount of synchronization required and the transmitted waveform.  The ``very low'' complexity attacks simply involve transmitting AWGN into a continuous band.  The ${J/S}_{F}$ metric is the overall J/S when taking into account the entire LTE frame, and is given by
\begin{equation} \label{jsf}
{J/S}_{F} =  {J/S}_{RE} \frac{ N_{RE}}{20 N_{symb} N_{sc} N_{RB}}
\end{equation}
where $N_{RE}$ is the number of resource elements associated with the physical channel or signal, $N_{RB}$ is the number of resource blocks used in the downlink or uplink signal, $N_{symb}$ is the number of OFDM symbols in a slot, and $N_{sc}$ is the number of subcarriers per resource block.  This calculation assumes a uniform power spectral density across the LTE downlink or uplink signal.  From the perspective of the jammer, a lower ${J/S}_{F}$ is better.

\section{Conclusion}
In this paper we analyzed the vulnerability of LTE to jamming by investigating the various physical channels and signals within LTE.  Using barrage jamming as a baseline, we have shown that much more effective jamming methods can be realized by exploiting the protocols of LTE.  In order to compare several methods, we derived metrics related to effectiveness and complexity for each one. When considering how many forms of jamming are more effective than barrage jamming, it is clear that LTE is extremely vulnerable to adversarial jamming.  In particular, the PCFICH and PUCCH are weak points in the downlink and uplink signal respectively.  This is not a surprising result, considering LTE was not designed to be a military communication system.  However, with the rapid growth of mobile devices, LTE is going to be a highly relied upon technology.

\bibliographystyle{IEEEtran}
\bibliography{references}

\end{document}